\documentclass{aastex}
\usepackage{spr-astr-addons}
\usepackage{url}
\usepackage{graphics}
\usepackage{mathptmx}
\usepackage{natbib}

\RequirePackage{color}

\newcommand{\teff}{$T_{\!\mbox{\scriptsize\em eff}}$}

 \newcommand{\hii}{\ion{H}{2}}

\begin{document}

\title{Distances to Galaxies from the Brightest Stars in the Universe}
\slugcomment{Astrophysics and Space Sciences}
\shorttitle{Distances from the Brightest Stars}
\shortauthors{R-P. Kudritzki}

\author{Rolf-Peter Kudritzki\altaffilmark{1,2}} \and \author{Miguel A. Urbaneja}
\affil{Institute for Astronomy, University of Hawaii, 2680 Woodlawn Drive, Honolulu, HI 96822}


\altaffiltext{1}{Max-Planck-Institute for Astrophysics, Karl-Schwarzschild-Str.1, D-85741 Garching, Germany}
\altaffiltext{2}{University Observatory Munich, Scheinerstr. 1, D-81679 Munich, Germany}

\begin{abstract}

Blue Supergiants (BSGs) are the brightest stars in the universe at visual 
light with absolute magnitudes up to M$_{V}$ = -10 mag. They are ideal stellar 
objects for the determination of extragalactic distances, in particular, 
because the perennial uncertainties troubling most of the other stellar 
distance indicators, interstellar extinction and metallicity, do not 
affect them. The quantitative spectral analysis of low resolution spectra 
of individual BSGs provides accurate stellar parameters and chemical 
composition, which are then used to determine accurate reddening and 
extinction from photometry for each individual object. Accurate distances 
can be determined from stellar gravities and effective temperatures 
using the "Flux Weighted Gravity - Luminosity Relationship (FGLR)". 

Most recent results of the quantitative spectral analysis of BSGs in 
galaxies within and beyond the Local Group based on medium and low 
resolution spectra obtained with the ESO VLT and the Keck telescopes on
Mauna Kea are presented and distances obtained with the FGLR-method 
are discussed together with the effects of patchy extinction and 
abundance gradients in galaxies. BSG metallicities and metallicity 
gradients are compared with results from strong-line \hii~region studies 
and the consequences for the empirical calibration of the metallicity 
dependence of the Cepheid period - luminosity relationship are pointed 
out. The perspectives of future work are discussed, the use of the
giant ground-based telescopes of the next generation such as the TMT on
Mauna Kea and the E-ELT and the tremendous value of the GAIA
mission to allow for the ultimate calibration of the FGLR using 
galactic BSGs.

\end{abstract}



\section{Introduction}

After the detection of the accelerated expansion of the universe the physical explanation of dark energy
has become the major challenge of astronomy and physics. One way to constrain the physics behind dark energy 
is to measure the equation-of-state parameter w = p/($\rho$c$^{2}$).
This requires an extremely accurate determination of the extragalactic distance scale and the Hubble Constant H$_{0}$.
As is well known \citep{macri06}, the determination 
of cosmological parameters from the cosmic microwave background is affected by degeneracies in parameter 
space and cannot provide strong constraints on the value of H$_{0}$ \citep{spergel06, tegmark04}.
Only if additional assumptions are made, for instance that the universe is flat, H$_{0}$ can be predicted with high 
precision (i.e. 2\%) from the observations of the cosmic microwave background, baryonic acoustic 
oscillations and type I high redshift supernovae. If these assumptions are relaxed, then much larger 
uncertainties are introduced \citep{spergel07, komatsu09}. The uncertainty of the 
determination of w is related to the uncertainty of H$_{0}$ through $\Delta$w/w $\approx$ 2$\Delta$H$_{0}$/H$_{0}$.
Thus, an independent determination of H$_{0}$ with an accuracy of 5\% will allow the uncertainty of w to 
be reduced to $\pm$0.1 or, even more ambitious, H$_{0}$ accurate to 1\%
will yield w accurate to 0.02.

Extremely promising steps towards this goal have been made recently by \citet{macri06} and by  
\citet{riess09a,riess09b,riess11}, who in a most ambitious approach use the maser galaxy NGC 4258 
and its Cepheids as a new anchor point for the extragalactic distance scale. HST detections of Cepheids out to 30 Mpc 
in galaxies with SNIa, which were detected recently and have accurate and well understood light cuves, are then used 
to consistently calibrate SNIa as far reaching standard candles using HST H-band photometry of the host galaxy Cepheids.
In this way, \citet{riess11} have been able to determine H$_{0}$ with unprecedented accuracy. They estimate an error of 
only 4\% including both systematic and random uncertainties (see papers by Lucas Macri and Adam Riess in this volume).

However, it is clear that the complexity of this bold and courageous approach
requires additional and independent tests. There are two long-standing issues with the use of Cepheids as
stellar distance indicators, the perennial problem of interstellar extinction and the uncertainty about a potential 
metallicity dependence of the period-luminosity relationship. While both problems can be mitigated to some extend 
by near IR observations and a strictly differential approach, as applied by \citet{riess11}, at a few percent accuracy 
level of H$_{0}$ they continue to be a major concern. 

In order to address these concerns, an independent and complementary method is desirable, which can overcome the 
problems of interstellar extinction and variations of chemical composition. In this paper, we introduce such a method,
the flux weighted gravity - luminosity relationship (FGLR), which is based on the quantitative spectroscopy of
blue supergiant stars (BSGs).

In the following we briefly illustrate the still acute problems caused by extinction and metallicity. We then introduce 
the spectroscopy of BSGs as an ideal tool to obtain accurate information about reddening, metallicity 
and distance. We will discuss most recent applications of the BSG-method and future work.

\section{The perennial problem of interstellar extinction}

\begin{figure}[!]

\begin{center}
\includegraphics[width=5cm, angle=90]{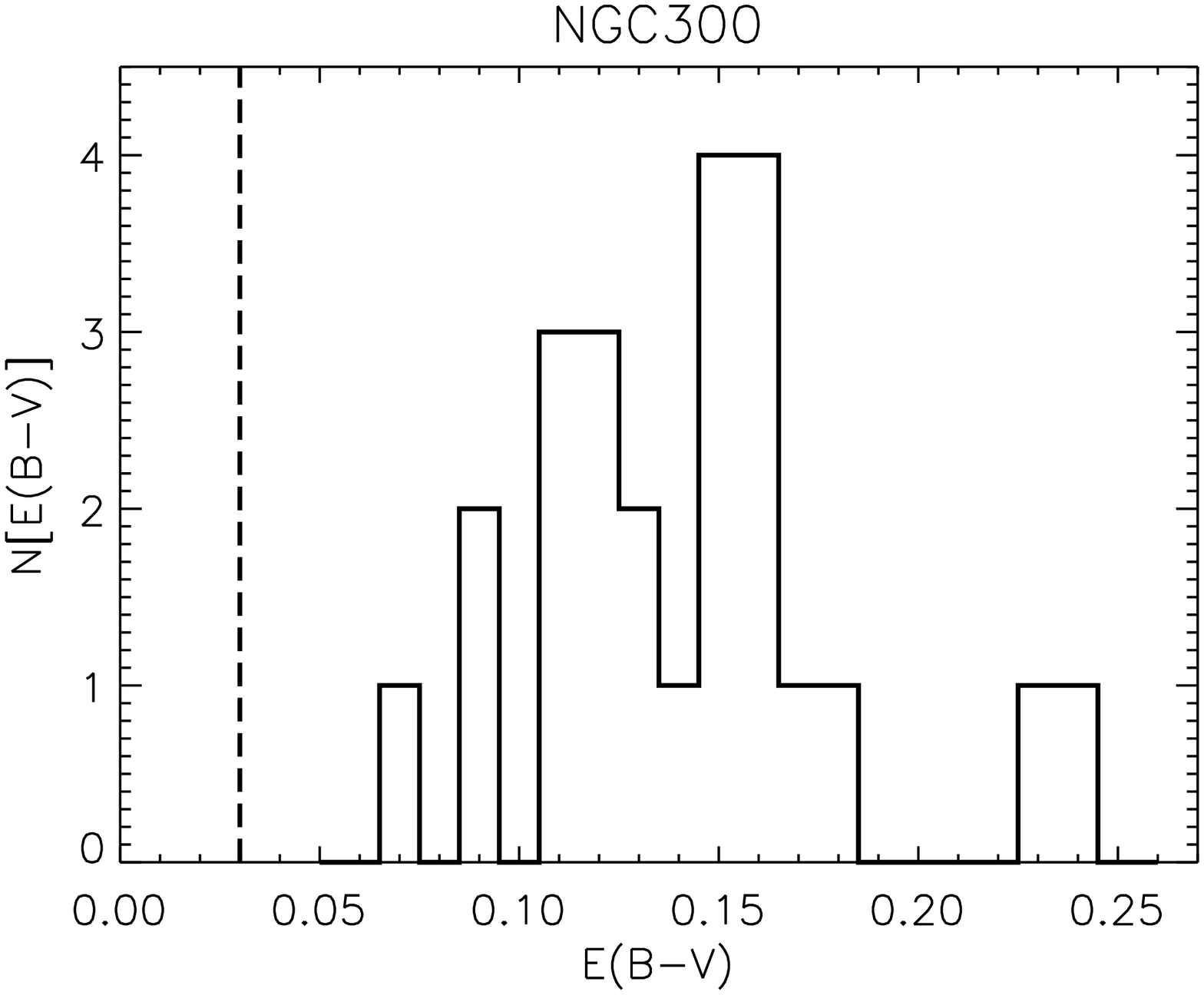} 
\includegraphics[width=5cm, angle=90]{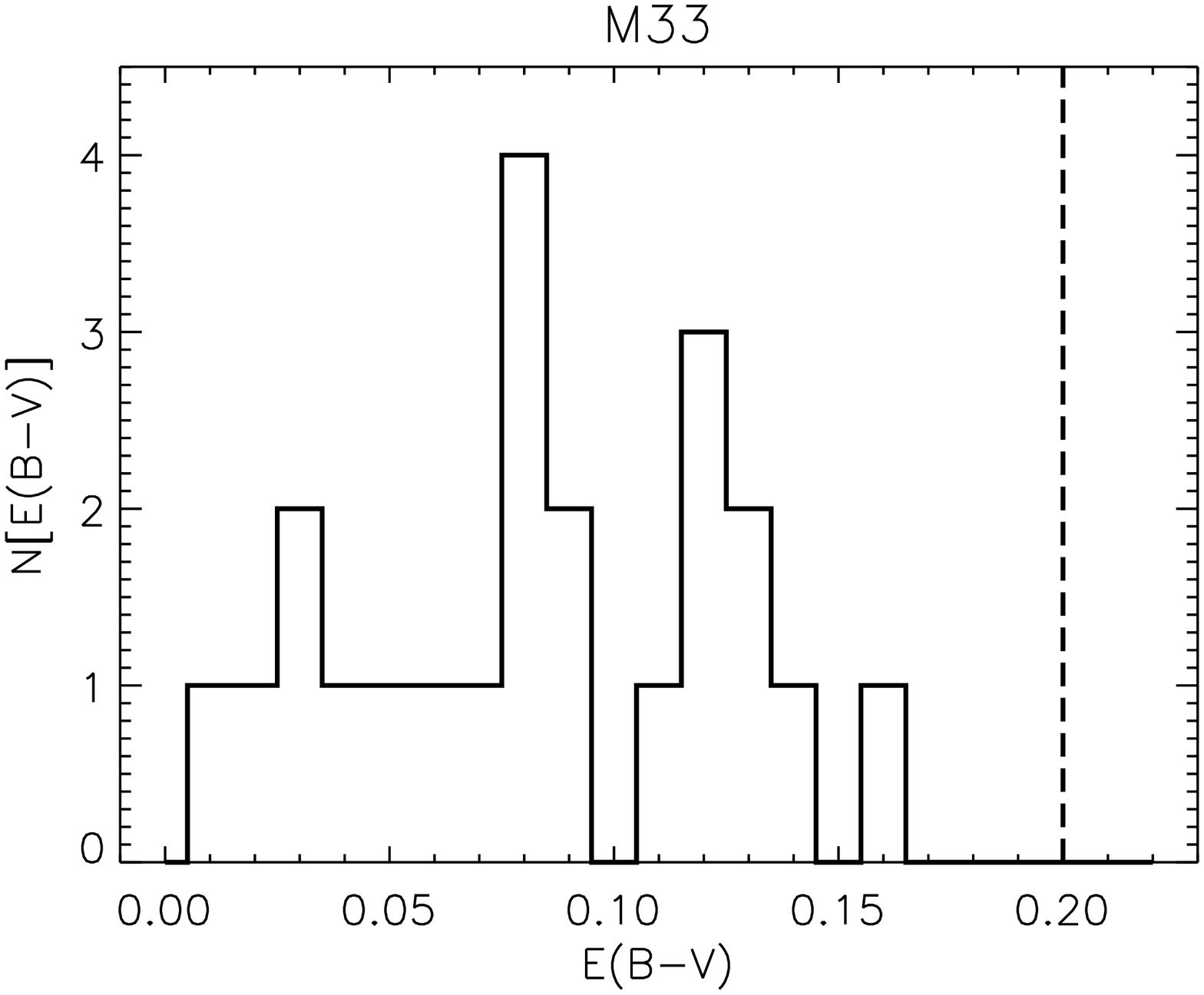} 
\includegraphics[width=5cm, angle=90]{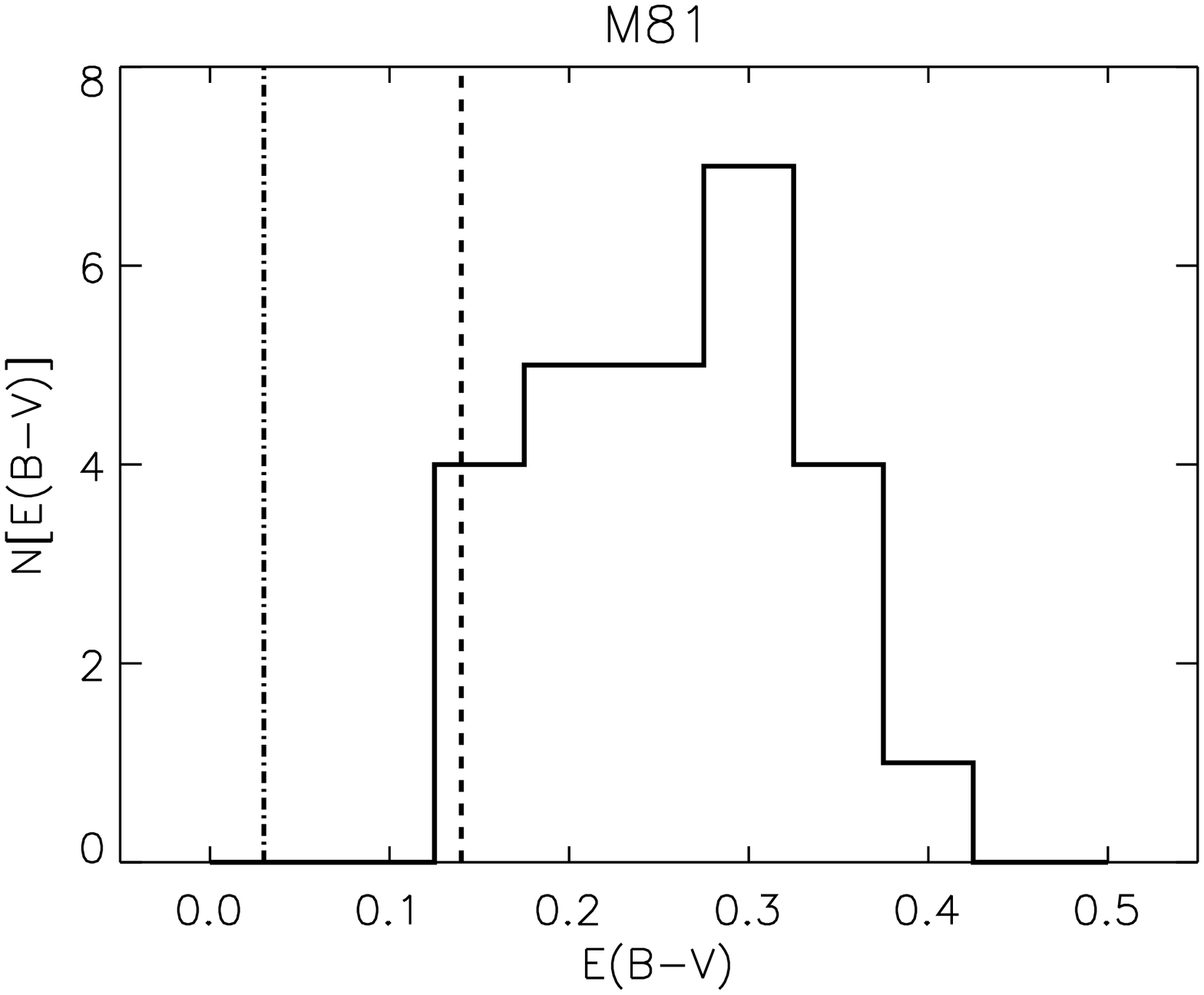} 

\caption{\small Histogram of the reddening distribution of BSGs in three galaxies determined from spectral analysis and photometry and compared
with the values adopted by the {\em HST} Key Project \citet{freedman01} (dashed). The dashed-dotted line represents the value from an
earlier KP paper \citep{freedman94}. BSG data for NGC 300 and M81 from \citet{kud08,kud11} and for M33 from \citet{u09}.}
   \label{fig1}
\end{center}
\end{figure}

Among the remaining uncertainties affecting the extragalactic distance scale, probably the most important one is 
interstellar reddening. As young stars, Cepheids tend to be embedded in dusty regions which produce a significant 
internal extinction, in addition to the galactic foreground extinction. Since Cepheids and BSGs belong to the same 
population and have a similar age, we can use the observations of BSGs to test whether reddening estimates for 
Cepheids seem reliable or not. Fig.\,\ref{fig1} shows the reddening distributions of BSGs
in three galaxies, NGC~300, M33 and M81, compared with the reddening value obtained in the {\em HST} Key Project (KP) \citep{freedman01}.
We note that the reddening in these star forming spirals is extremely patchy and leads to a wide distribution of E(B-V).
Moreover, the average reddening values obtain by the KP are in disagreement with the observed BSG reddening distribution.
For NGC~300 and M81, reddening is underestimated and the opposite is the case for M33. This is alarming, since reddening errors 
of 0.1 to 0.2 mag transform into errors of 0.3 to 0.6 mag in the extinction corrected distance modulus.

\begin{figure}[h]

\begin{center}
  \includegraphics[width=8cm]{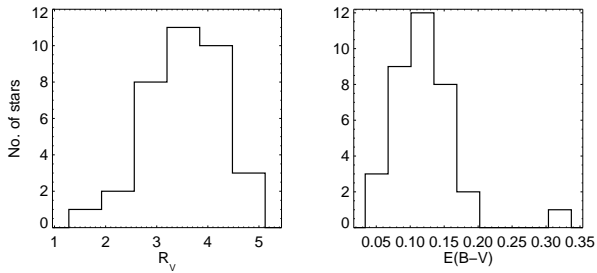} 

\caption{\small Histograms of the reddening  distribution of BSGs in the LMC. Left: R$_{V}$; Right: E(B-V). 
Data from \citet{urbaneja11}.}
   \label{fig2}
\end{center}
\end{figure}

\begin{figure}[h]

\begin{center}
 \includegraphics[width=0.3\textwidth,angle=90]{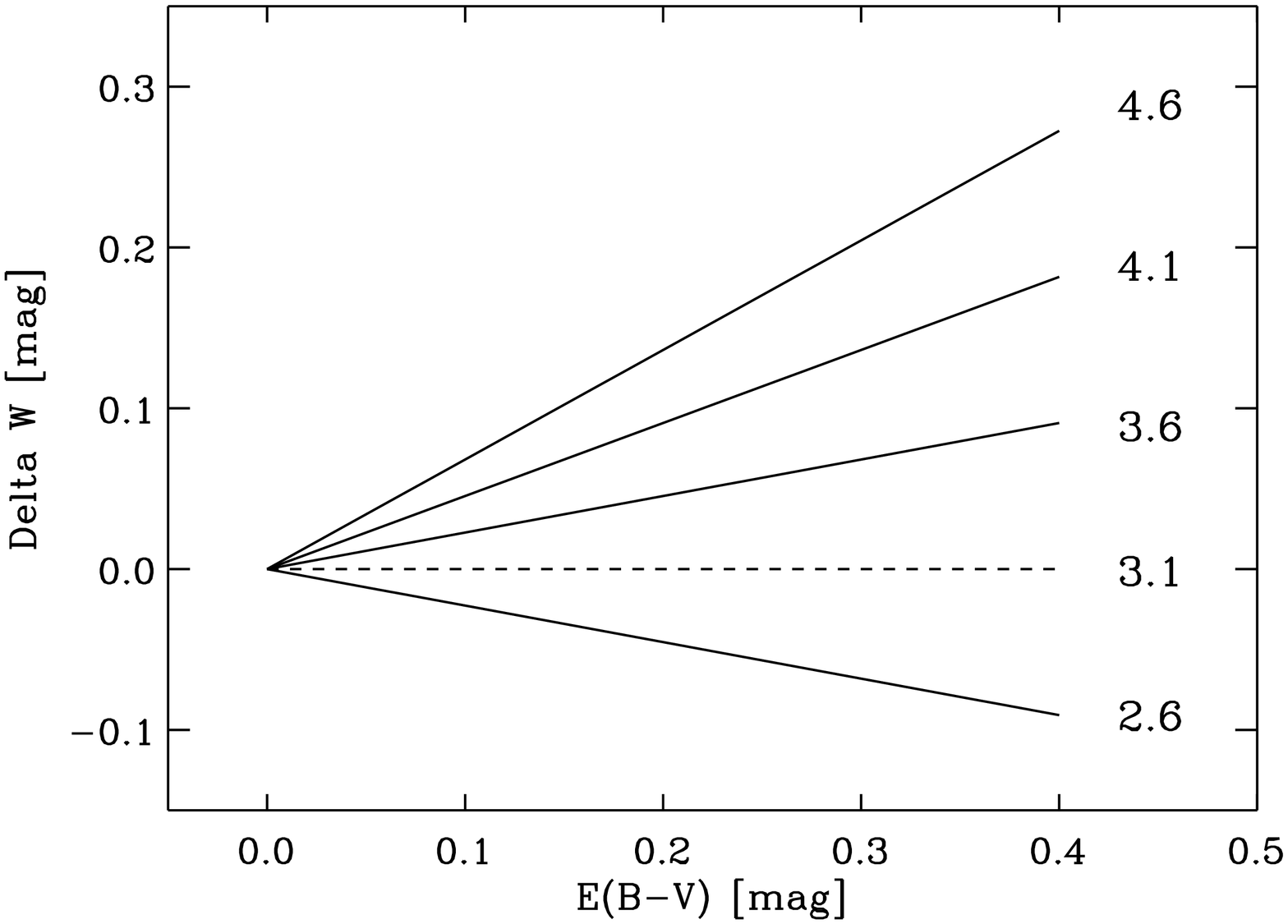} 
\includegraphics[width=0.3\textwidth,angle=90]{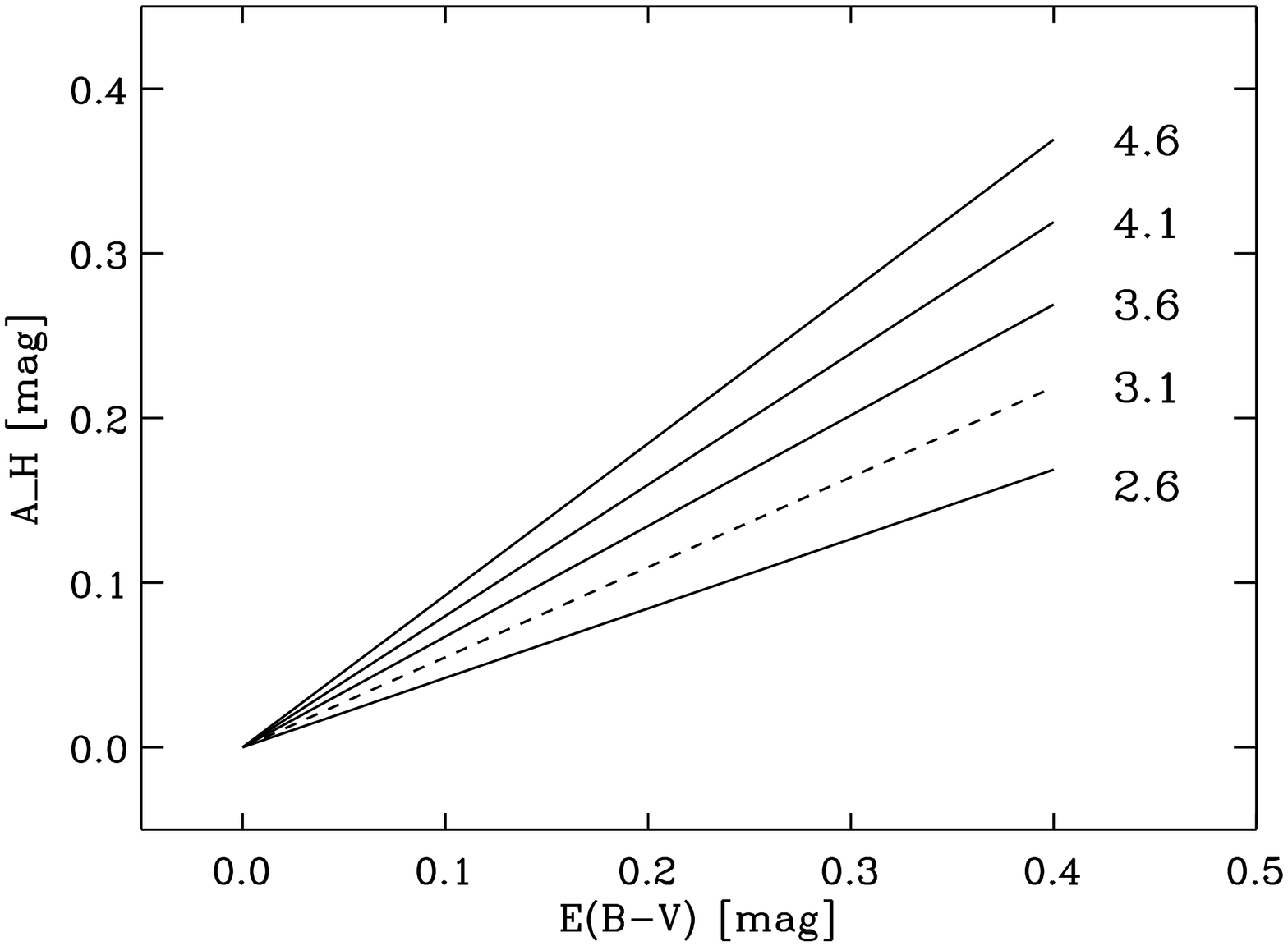} 

\caption{\small Change in \textquotedblleft reddening-free\textquotedblright Wesenheit-magnitude $\Delta$W (top panel) and extinction at H-band A$_{\rm H}$ (bottom panel) 
as a function of reddening E(B-V) for different values R$_V$.}
   \label{fig3}
\end{center}
\end{figure}

In order to avoid distance moduli errors due to reddening the KP and most of the more recent Cepheid work, 
which used optical photometry (B, V, I), has applied the Wesenheit method, where a ``reddening-free'' 
combination of Cepheid V and I magnitudes, W=V-R$_{VI}$(V-I) with R$_{VI}$=A$_{V}$/(A$_{V}$-A$_{I}$) 
is used for the fit of the period-luminosity relationship (PLR) \citep{gieren04, macri06, mccommas09, shappee11, gerke11}.
A$_{V}$ and A$_{I}$ are the extinction at V- and I-band, respectively, and the ratio R$_{VI}$=2.45 is  obtained 
from \citet{schlegel98}, who used the MW ``standard reddening-law'' in the analytical formulation given 
by \citet{cardelli89} for R$_{V}$ = 3.1. After the determination of the distance, the fit of the PLRs 
in V and I band allow for an estimate of average reddening.
As illustrated by Fig.\,\ref{fig1} this approach does not always lead to good agreement with the 
reddening information obtained from BSGs, although in some cases good agreement is found (compare \citealt{gieren04} and 
\citealt{kud08}).

It is important to remember that in star forming galaxies large deviations from the standard value R$_{V}$ = 3.1 
are frequently encountered. Fig.\,\ref{fig2} shows the result of a very recent BSGs study in the LMC. How would 
such deviations affect the ``reddening-free'' W-magnitude as defined above?  Fig.\,\ref{fig3} demonstrates that 
the effects can be alarmingly large and certainly have the potential to influence the determination of distance moduli.
The shift of W-magnitude $\Delta$W caused by a change of R$_{V}$ is well descibed by $\Delta$W=0.454(R$_{V}$-3.1)*E(B-V).

Reddening effects become less important at near-IR wavelengths. In the Araucaria-Project, the combination of B, V, I 
photometry with J- and K-band observations has been used very successfully (see, for instance, \citealt{gieren05}). In a similar way,
\citet{riess09b, riess11} have used H-band photometry with HST. However, while extinction is much smaller in the near-IR, 
it is not zero, as illustrated by Fig.\,\ref{fig3}. For ``normal'' reddening with R$_{V}$ = 3.1 an error of $\Delta$E(B-V)=0.2 mag
leads to an extinction error $\Delta$A$_{H}$=0.12 mag equivalent to a distance error $\Delta$d/d=5.5\%. Equivalently, a reddening 
law with R$_{V}$=4.6 at E(B-V)=0.3 mag produces a similar effect.Obviously, with the 
ambitious goal determine the H$_{0}$ accurate to 1\% this is something to keep in mind. This means that
entirely independent and complementary approach to address the issue of reddening is highly desirable.

\begin{figure}[h]

\begin{center}
\includegraphics[width=6cm]{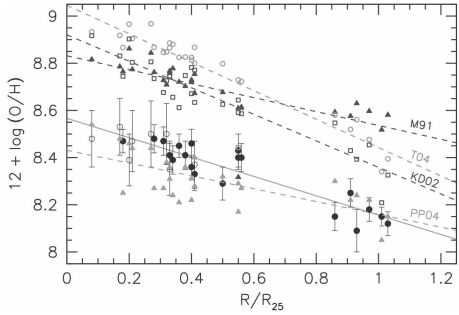} 

\caption{\small The mass-metallicity relationship of star forming galaxies in the nearby 
universe obtained by applying several widely used empirical metallicity calibrations based on 
different strong line ratios. This figure illustrates that there is an effect not only on the 
absolute scale, but also on the relative shape of this relationship. 
Adapted from \citet{kewley08}.}
   \label{fig4}
\end{center}
\end{figure}

\begin{figure}[h]

\begin{center}
\includegraphics[width=7cm]{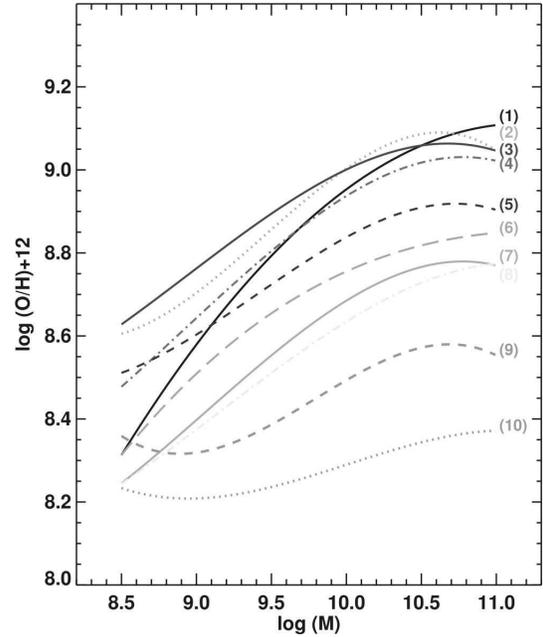} 

\caption{\small \hii~region galactocentric oxygen abundance gradients in NGC~300 obtained from 
our dataset but different strong line  calibrations: 
 McGaugh~1991\,=\,M91, Tremonti et al.~2004\,=\,T04, Kewley \& Dopita~2002\,=\,KD02, 
and Pettini \& Pagel~2004\,=\,PP04, as shown by the labels to the corresponding 
least squares fits. The auroral line-based abundances determined 
by Bresolin et al.~(2009) are shown by the full and open circle symbols, and 
the corresponding linear fit is shown by the continuous line.
}
   \label{fig5}
\end{center}
\end{figure}

\section{The important role of metallicity}

An equally important uncertainty in the use of Cepheids as distance indicators is the dependence of the PLR on metallicity. 
Comparing HST photometry of Cepheids in inner and outer galactic fields \citet{kennicutt98} (for M101) and \citet{macri06}
(for the maser galaxy NGC 4258) have found that inner field Cepheids yield shorter distance moduli by about 0.2 mag than the ones in 
the outer fields. They conclude that this is caused by the metallicity gradient of spiral galaxies and 
that {\it the brightness of Cepheids increases with metallicity}. Using metallicity information from \hii~region oxygen 
emission lines they adopt a metallicity correction of the distance modulus $\mu$ of the form $\Delta\mu$=$\gamma$([Z]-[Z]$_{LMC}$),
where [Z]=logZ/Z$_{\odot}$ is the metallicity relative to the sun on a logarithmic scale. $\gamma$-values between -0.2 and -0.3 were found.
This agrees with \citet{sakai04}, who related the difference between TRGB and Cepheid distances to the galactic \hii~region metallicities 
(not taking into account metallicity gradients, though) and derived a similar PLR dependence on metallicity. 

However, these results are highly uncertain. \citet{rizzi07} have argued that many of the TRGB distances used 
by \citet{sakai04} need to be revised. With the \citet{rizzi07} TRGB distances the \citet{sakai04} 
dependence of the PLR on metallicity disappears and the results are in much closer agreement with 
stellar pulsation theory \citep{fiorentino02, fiorentino07, marconi05, bono08}, which predicts just the opposite, namely that {\it the brightness 
of Cepheid decreases with metallicity}. 

It is very important to note that all the \hii~region (oxygen) metallicities adopted, when 
comparing Cepheid distances with TRGB distances or when using metallicity gradients, are highly uncertain. They 
result from the application of a simplified analysis method, the so called ``strong-line method'', which uses only the fluxes 
of the strongest forbidden lines of (most commonly) [OII] and [OIII] relative to H$_{\beta}$. Unfortunately, abundances 
obtained with the strong-line method depend heavily on the calibration used. As a striking example, \citet{kewley08}
have demonstrated that the quantitative shape of the mass-metallicity relationship of galaxies can change from very steep 
to almost flat depending on the calibration used (Fig.\,\ref{fig4}). In the same way, as shown by \citet{kud08} 
and \citet{bresolin09}, metallicity gradients of spiral galaxies can change from steep to flat and 
absolute values of metallicity can shift by as much as 0.6 dex, again as the result of different calibrations of the 
strong line method (Fig.\,\ref{fig5}). The Cepheid work on spiral galaxies usually uses the calibration by \citet{zaritsky94}.
As shown by \citet{kud08}, this calibration gives far too high metallicities.

The results displayed in Fig.\,\ref{fig4} and \ref{fig5} are shocking. They reveal that at 
present galaxy metallicities based on strong-line methods are uncertain by 0.6 to 0.8 dex because of the systematic 
uncertainties inherent in the strong line methods used. Even with a strictly differential approach in the use 
of Cepheids as distance indicators as applied by \citet{riess11} this still creates residual uncertainties, which become 
important in view of the ambitious goal to determine H$_{0}$ with an accuracy of a few percent.
This major problem requires a fresh approach and is begging 
for the development of  a new and independent method less affected by systematic uncertainties.
Very obviously, in order to better assess the systematic uncertainties of the determination of distances to star-forming spiral and 
irregular galaxies in the local universe an independent and complementary method is desirable, which can overcome 
the problems of interstellar extinction and variations of chemical composition. In the next section we introduce 
such a method, which is based on the quantitative spectroscopy of BSGs. The method makes use 
of the enormous intrinsic brightness of these objects.

\begin{figure*}
 \begin{center}
 \includegraphics[width=12cm]{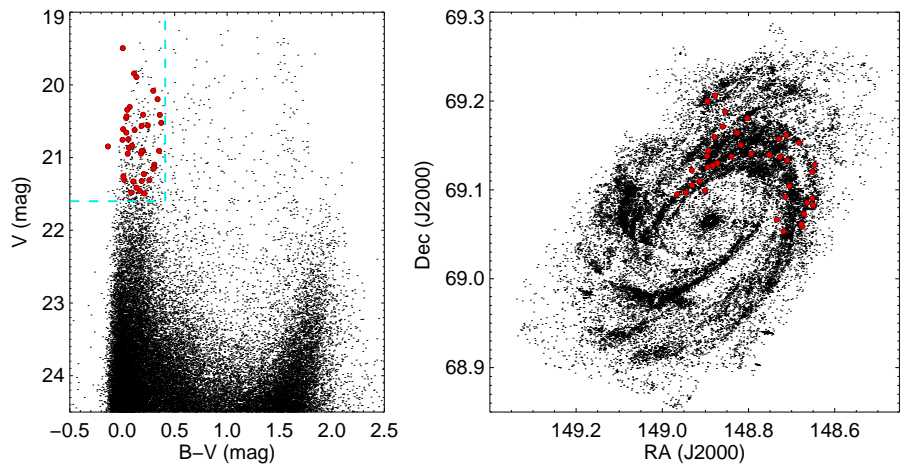}
  \caption[]{\small Selection of M81 BSG targets. Left: Color magnitude diagram (photometry from \citealt{dalcanton09}) 
with selection box (blue dashed) and selected targets (red). Right: Location of selected targets within M81. From \citet{kud11}.}
  \label{fig6}
 \end{center}
\end{figure*}

\begin{figure}

\begin{center}
\includegraphics[width=7cm]{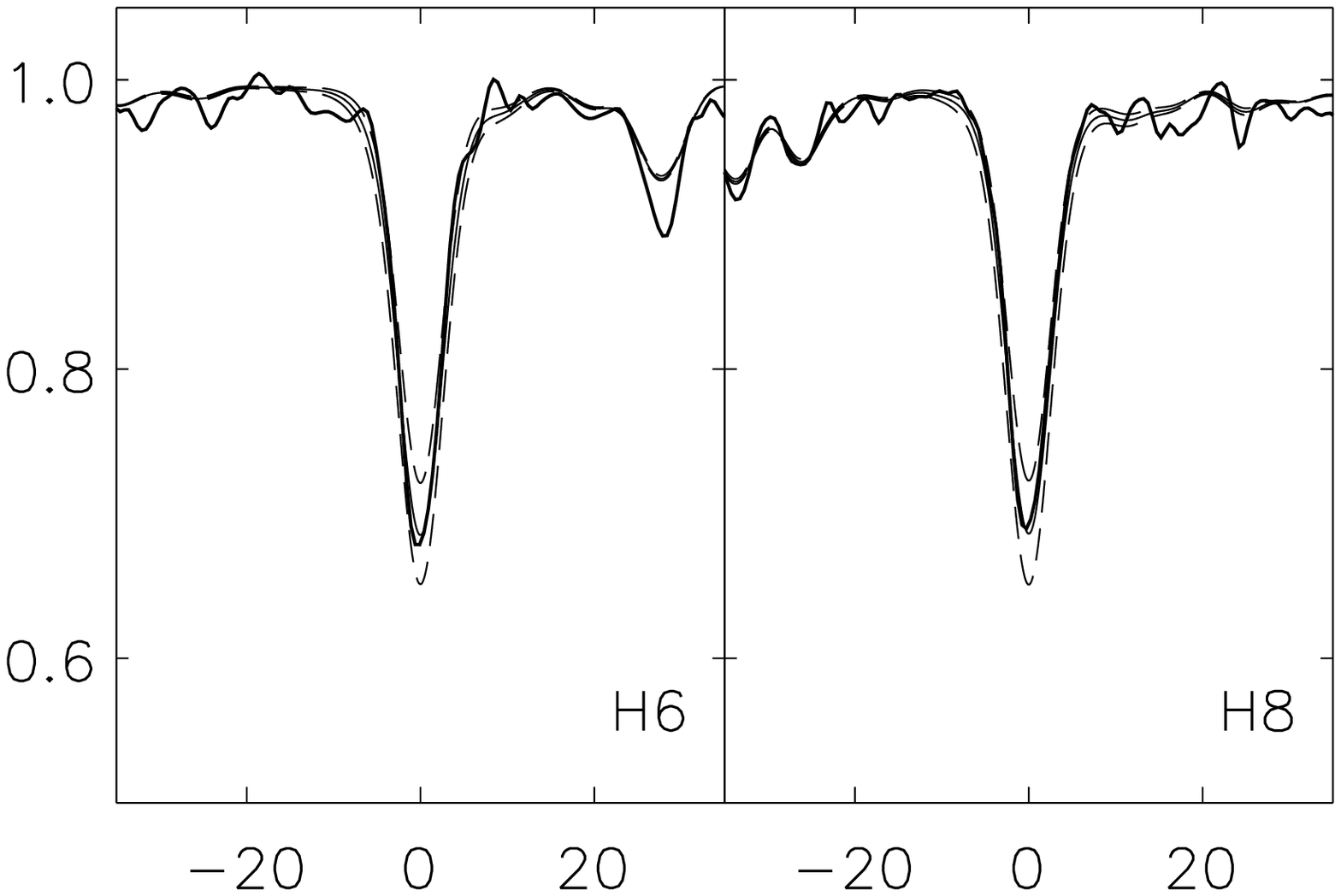}
\includegraphics[width=6cm, angle=90]{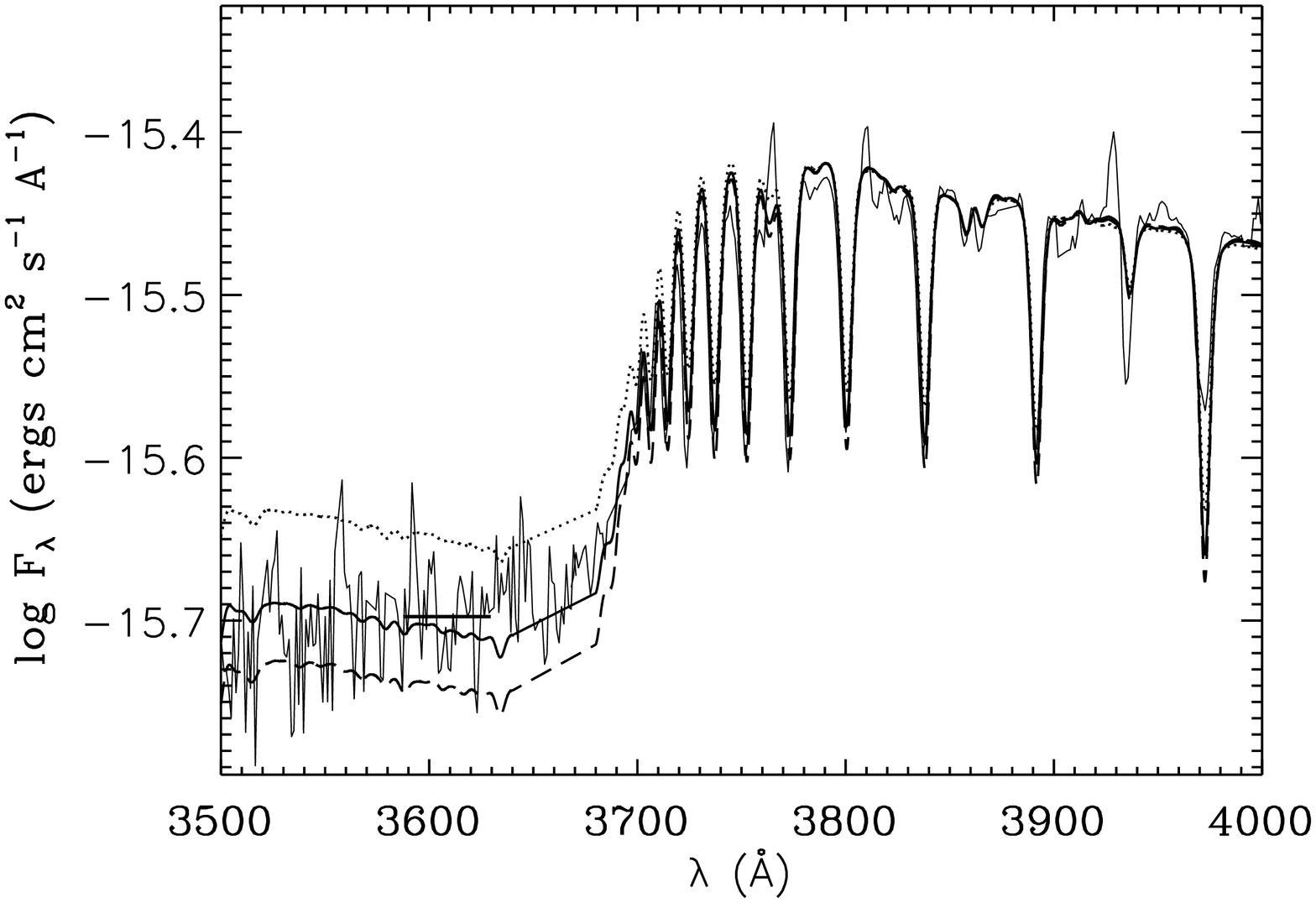}

\caption{\small From \citet{kud08}: Model atmosphere fit of NGC~300 No. 21. Top: Two observed Balmer lines of 
fitted with \teff\/ = 10000~K and $log~g$ = 1.55 (solid) and $log~g$ = 1.45 and 1.65, respectively (dashed). The Balmer lines are used to determine log g.
Bottom:  Fit of the observed Balmer jump with \teff\/ = 10000~K and $log~g$ = 1.55 (solid). Two additional models with the same $log~g$
but \teff\/ = 9750~K (dashed) and 10500~K (dotted) are also shown. The balmer jump is used to determine \teff\/.}
   \label{fig7}
\end{center}
\end{figure}

\section{Blue supergiants come to rescue}

An obvious alternative method to constrain metallicity is the detailed quantitative 
spectroscopic analysis of individual stars in galaxies. While this may sound unfeasible at first glance, 
it is well within the capabilities of current technology. The brightest stars in the universe at visual light, 
blue supergiants (BSGs) of spectral type A and B are the perfect choice for this purpose. BSGs are massive stars 
in the mass range between 12  to 40 M$_\odot$ in the short-lived evolutionary phase (10$^{3}$ to 10$^{5}$ years) when they 
leave the hydrogen main sequence and cross the HR-diagram at constant luminosity and almost constant mass to 
become red supergiants. Because of Wien's law massive stars increase their brightness in visual light 
dramatically when evolving towards lower temperatures and reach absolute visual magnitudes up to 
M$_{V} \approx$  -9.5 mag in the BSG phase, rivaling with the integrated light of globular clusters and 
dwarf galaxies. Because of their extreme brightness they are ideal tools to accurately determine the chemical 
composition of young stellar populations in galaxies. 

BSG spectra are rich in metal absorption lines from several elements (C, N, O, Mg, Al, S, Si, Ti, Fe, among others). 
As young objects  with ages of 10 Myrs they provide important probes of the current composition of the interstellar 
medium. Abundance determinations of these objects can therefore be used to trace the present abundance patterns in 
galaxies, with the ultimate goal of recovering their chemical and dynamical evolution history. In addition to the 
$\alpha$-elements the spectral analysis of supergiants provides the only accurate way to obtain information about 
the spatial distribution of Fe-group element abundances in external star-forming galaxies.
As outlined above, beyond the Local Group most of the information about the chemical properties of spiral galaxies 
has so far been obtained through the study of \hii~regions emission lines using strong-line methods, which have huge 
systematic uncertainties arising from their calibrations. Direct stellar abundance studies of BSGs open a completely 
new and more accurate way to investigate the chemical evolution of galaxies and are free of such uncertainties.

\begin{figure}

\begin{center}
\includegraphics[width=8cm]{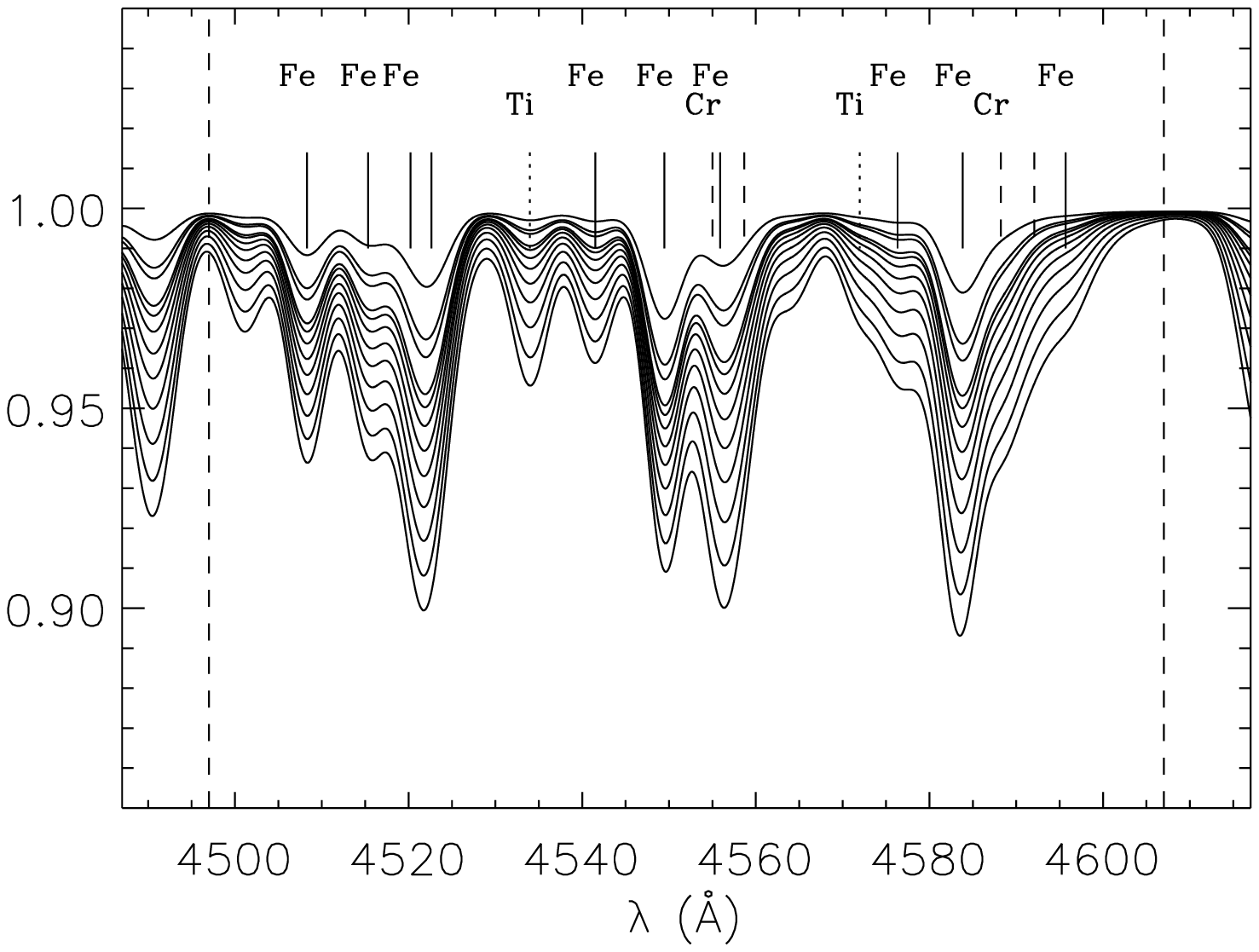} 

 \caption{\small Synthetic metal line spectra calculated for the stellar parameters of NGC~300 target 
No.21 as a function of metallicity in the spectral window from 4497~\AA~to 4607~\AA. 
Metallicities range from [Z] = -1.30 to 0.30 dex, as described in the text. The dashed 
vertical lines give the edges of the spectral window as used for a determination of 
metallicity. From \citet{kud08}.}
   \label{fig8}
\end{center}
\end{figure}

\begin{figure*}

\begin{center}
\includegraphics[width=12cm]{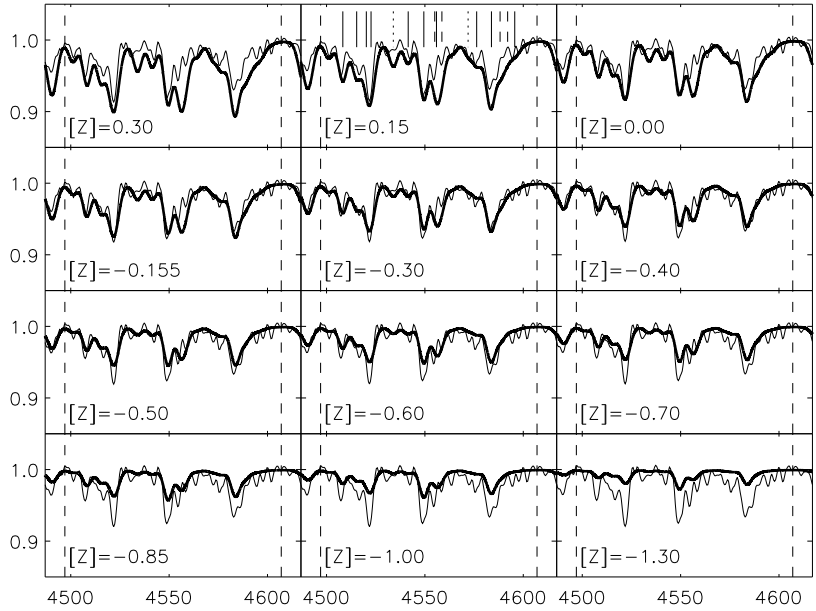} 

\caption{\small Observed spectrum of the same target as in Fig. 1 and 2 in the same spectral window as 
Fig.\,\ref{fig8} but now the synthetic spectra for each metallicity overplotted 
separately. From \citet{kud08}.}
   \label{fig9}
\end{center}
\end{figure*}

Based on detailed high resolution, very high signal-to-noise (S/N) studies of blue 
supergiants, which yield abundances as accurate as 0.05 dex \citep{przybilla06, schiller08, przybilla08},
\citet{kud08} developed an efficient new spectral diagnostic technique for
low resolution spectra (FWHM  $\sim 5$ \AA)  with good S/N ratio (50 or better), which allows for an
accurate determination of effective temperature, gravity, metallicity, interstellar reddening and
extinction. Metallicities accurate to 0.1 to 0.2 dex for each individual target can be obtained at this
lower resolution and S/N. The method has  been applied to irregular and spiral galaxies in the Local Group 
(WLM -- \citealt{bresolin06}; \citealt{urbaneja08}; NGC
3109 -- \citealt{evans07}; IC 1613 -- \citealt{bresolin07}; M33 -- \citealt{u09}) and 
beyond ( NGC~300-- \citealt{kud08}; M81-- \citealt{kud11}). In the following, we describe the diagnostic technique.

\begin{figure}

\begin{center}
 \includegraphics[width=8cm]{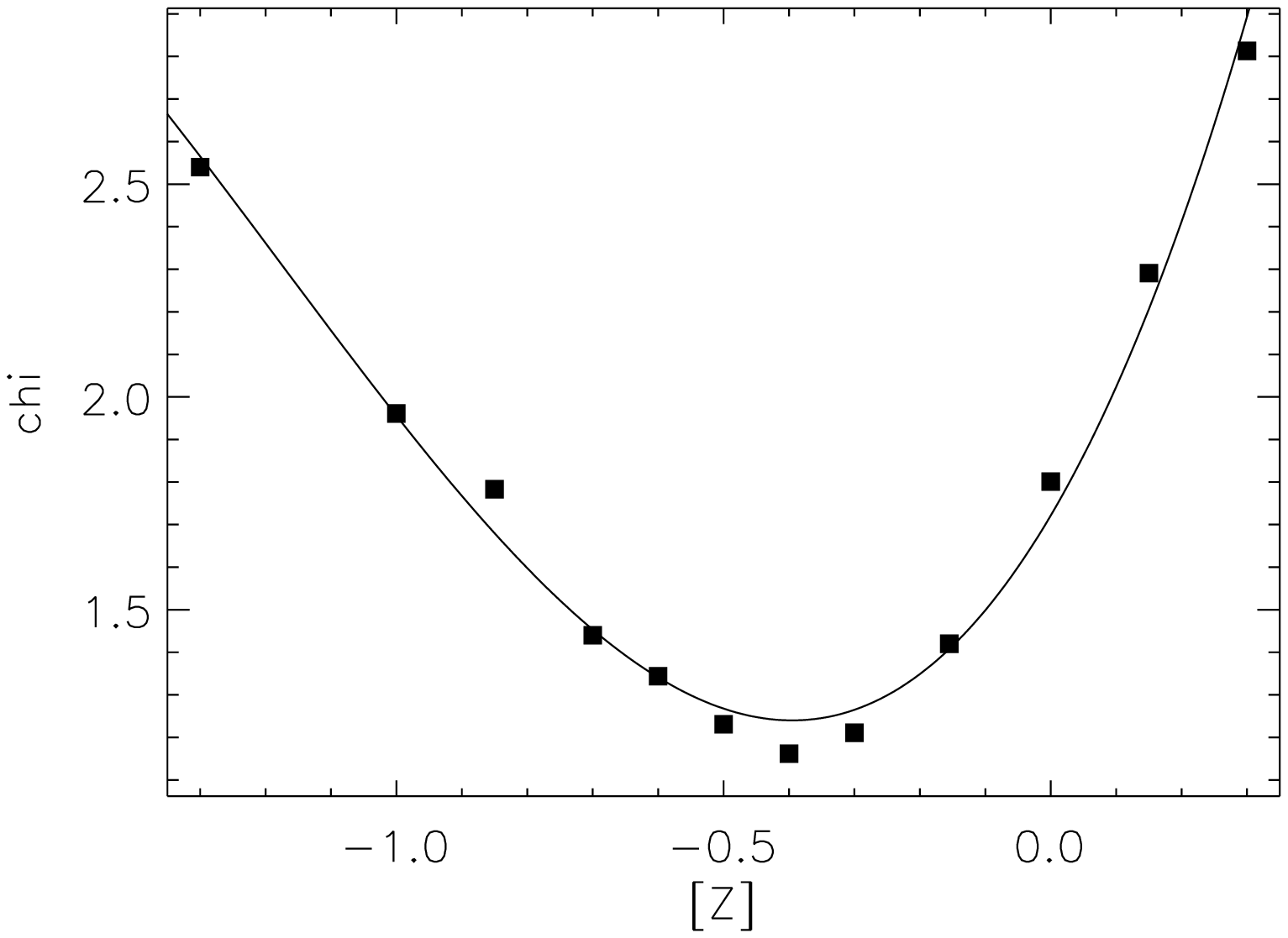} 

\caption{\small $\chi ([Z])$ as obtained from the comparison of observed and 
calculated spectra. The solid curve is a third order polynomial fit. From \citet{kud08}.}
   \label{fig10}
\end{center}
\end{figure}

Targets are selected from HST or ground-based photometry, usually B, V, I (Fig.\,\ref{fig6}). Spectra are then taken using MOS spectrographs 
such as FORS at the VLT or LRIS at Keck providing both flux calibrated and continuum rectified spectra. The effective 
temperature Teff is determined from the fit of the Balmer jump and gravity log g from the Balmer lines  (Fig.\,\ref{fig7}). For the 
values of  Teff and log g obtained we can then calculate synthetic spectra as a function of metallicity and 
compare them with the observed spectrum in several selected spectral windows. A $\chi^{2}$ -fit obtained from the 
comparison of observed and calculated spectra then yields the metallicity with an accuracy between 0.1 to 0.2 dex per object.
Fig.\,\ref{fig8}, \,\ref{fig9}, and \,\ref{fig10} illustrate the method. We also emphasize that with \teff\/, log g and [Z] 
determined we know the intrinsic SEDs and colors of the BSGs studied and, thus, can accurately determine reddening and 
extinction by comparing with the observed colors and SEDs.

\begin{figure}

\begin{center}
 \includegraphics[width=7cm]{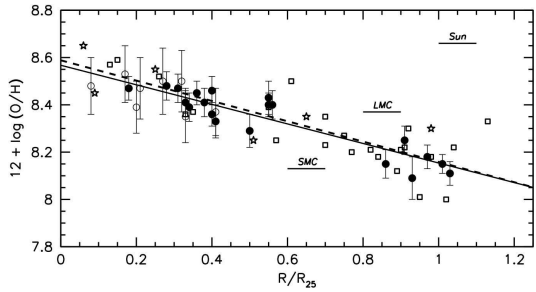} 

 \caption{\small Radial oxygen abundance  gradient obtained from \hii\ regions (circles) 
and blue supergiants (star symbols: B supergiants; open squares: A supergiants). The 
regression to the \hii\ region data is shown by the continuous line. The dashed line 
represents the regression to the BA supergiant star data. For reference, the oxygen 
abundances of the Magellanic Clouds (LMC, SMC) and the solar photosphere are 
marked. From \cite{bresolin09}.}
   \label{fig11}
\end{center}
\end{figure}

\begin{figure}

\begin{center}
 \includegraphics[width=6cm, angle=90]{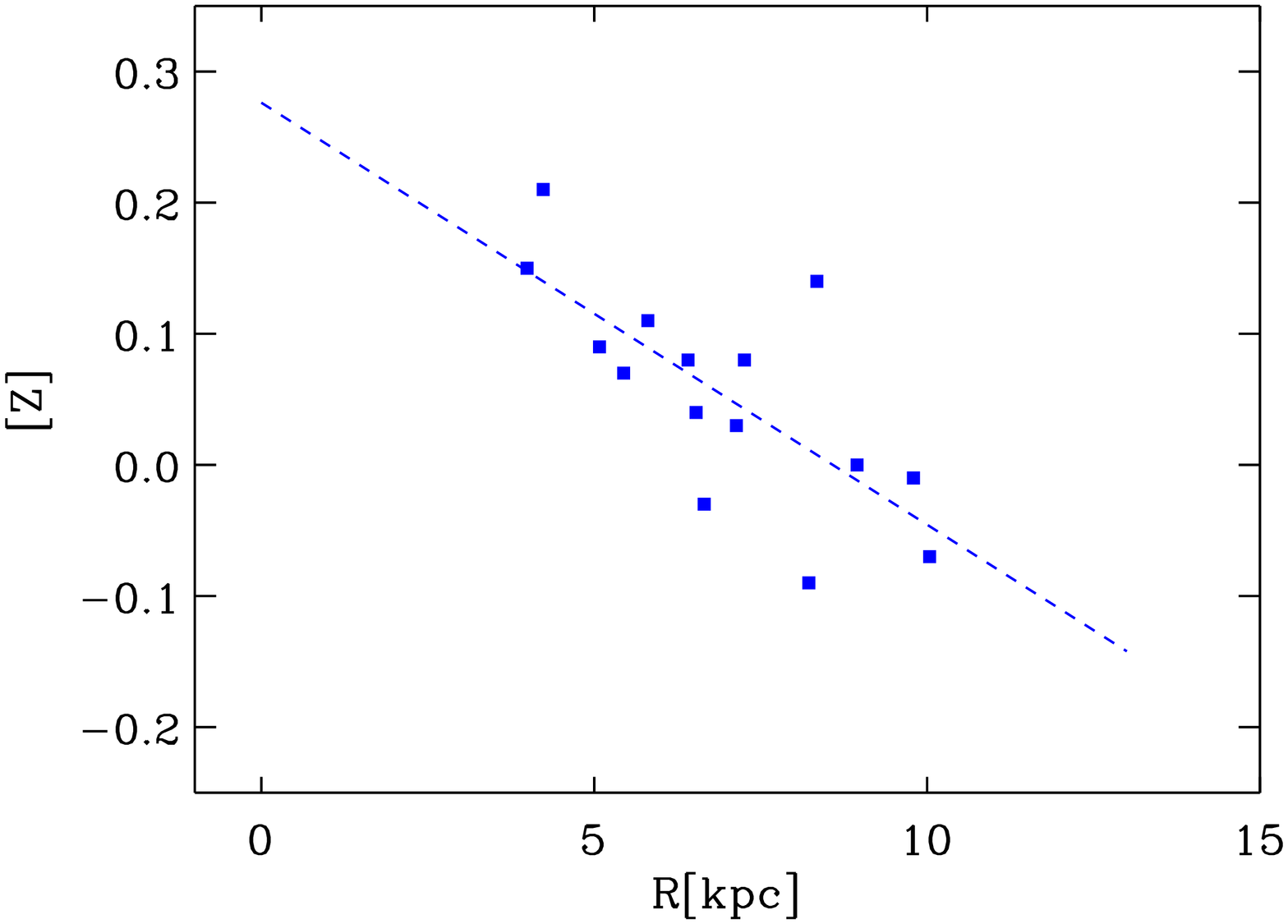} 

 \caption{\small Metallicity of BSGs in M81 as function of galactocentric distance in kpc.
          From \citet{kud11}.}
   \label{fig12}
\end{center}
\end{figure}

\citet{bresolin09} carried out the crucial test to compare blue supergiant 
with \hii~region metallicities in NGC 300, the latter obtained through very deep and long exposures revealing the weak 
nebular electron temperature sensitive auroral lines such as [OIII]$\lambda$4363. Auroral lines provide a classical and 
physically well-understood method to measure abundances of ionized nebulae, but its use requires a substantial 
additional observational effort at higher metallicity, where these lines are very weak. The excellent agreement between BSGs and the 
information obtained from \hii~regions using auroral lines is encouraging and promising (Fig.\,\ref{fig11}). 

\cite{bresolin09} also perform an important experiment. They use their new and very 
accurate measurements of strong line fluxes and simply apply a set of different strong line 
calibrations to obtain oxygen abundances without using the 
information from the auroral lines. The result of this experiment compared with the 
oxygen abundances using the auroral lines is shown in Fig.\,\ref{fig5}. As already discussed, the comparison 
is shocking. The abundance offsets introduced by the application 
of inappropriate strong-line calibrations can be as large as 0.6 dex, putting 
the whole business of defining metallicity corrections for Cepheid distance moduli through the measurement of nebular 
metallicities and metallicity gradients into jeopardy.

On the other hand, BSGs are an excellent tool to obtain accurate metallicity information about the young stellar 
population in galaxies. Fig.\,\ref{fig12} gives an example for the giant spiral M81 at 3.5 Mpc distance. This galaxy has 
slightly super-solar metallicity (0.2 dex) in the inner regions and a very shallow abundance gradient of 0.034 dex kpc$^{-1}$.
A linear regression yields [Z]=0.286-0.034R/kpc. We note that the \hii~region strong-line study by \citet{zaritsky94} 
obtains much higher metallicities and a slightly steeper gradient [Z]=0.51-0.041R/kpc, which is crucial for the Cepheid 
distance modulus metallicity correction (see section 7). (Here, oxygen is taken as a proxy for metallicity and we use 
[O/H]=12+log(O/H)=8.69 from \citealt{allende01} as the value for the solar oxygen abundance).

The BSG studies of the less massive galaxies NGC~300 \citep{kud08} and M33 \citep{u09} resulted in 
gradients of 0.08 and 0.07 dex kpc$^{-1}$, respectively, steeper than M81 but still very shallow. For the 
metallicity correction of Cepheid distance moduli this has severe consequences, as will be discussed in section 7. 

\begin{figure}

\begin{center}
 \includegraphics[width=6cm,angle=90]{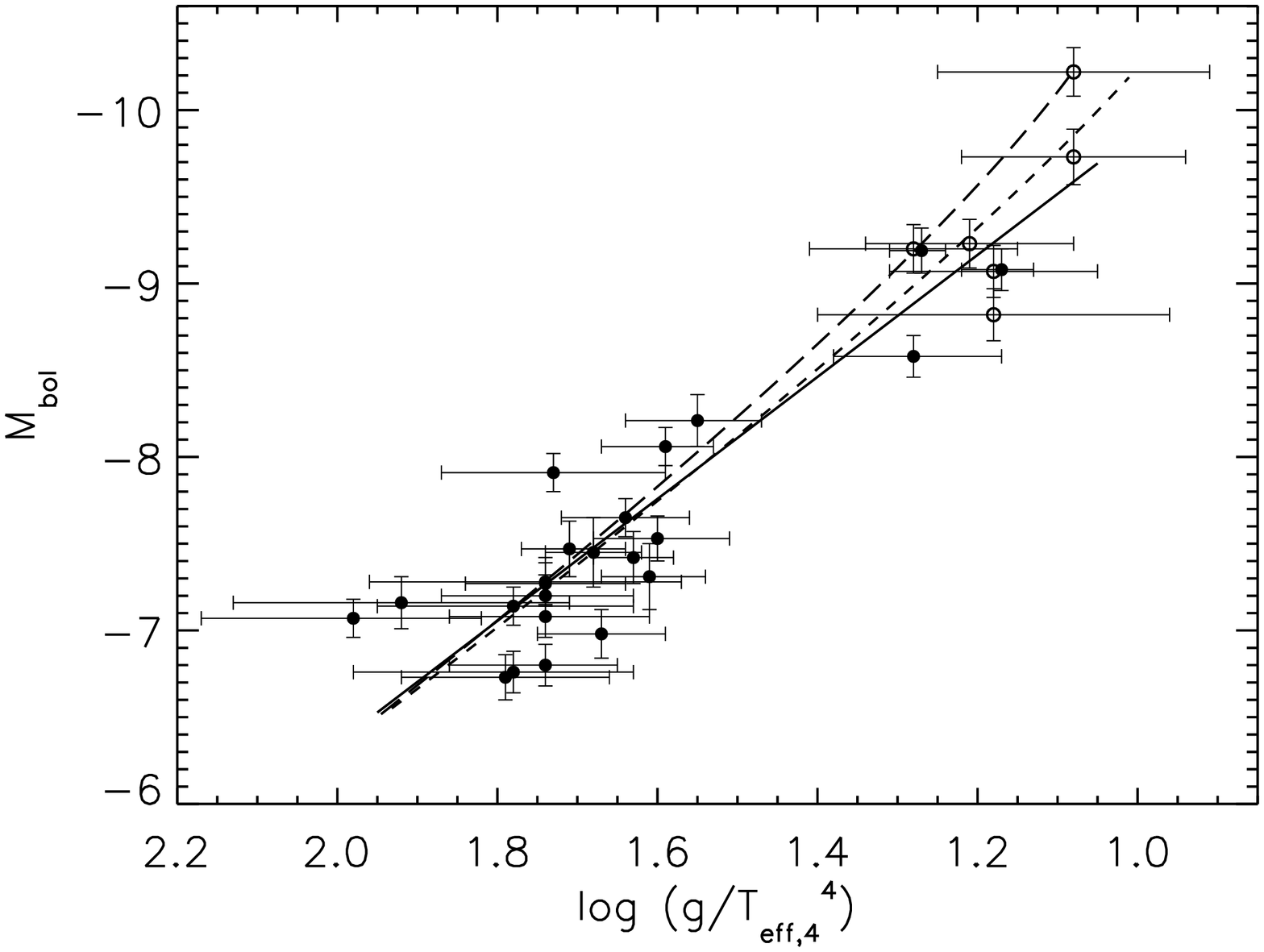} 

\caption{\small The FGLR of A (solid circles) and B (open circles) supergiants in 
NGC 300 and the linear regression (solid). The stellar evolution 
FGLRs for models with rotation are also overplotted 
(dashed: Milky Way metallicity, long-dashed: SMC metallicity). From \citet{kud08}.
}
   \label{fig13}
\end{center}
\end{figure}

\begin{figure*}

\begin{center}
 \includegraphics[width=10cm,angle=90]{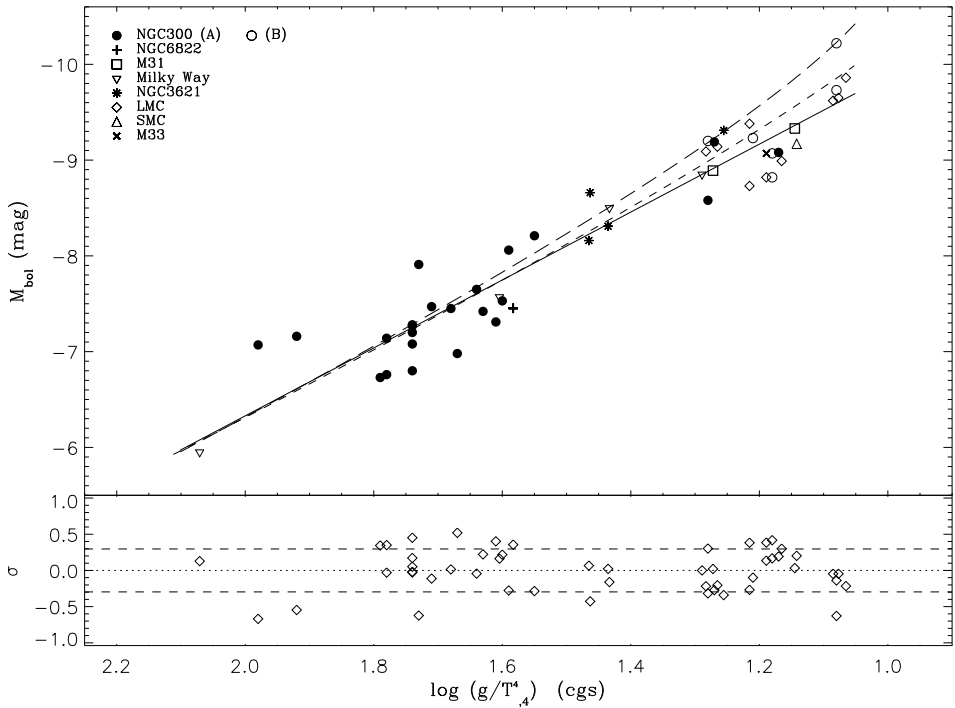} 

\caption{\small The FGLR of A (solid circles) and B (open circles) supergiants in 
8 galaxies including NGC 300 and the linear regression (solid). The stellar evolution 
FGLRs for models with rotation are again overplotted. From \citet{kud08}.
}
   \label{fig14}
\end{center}
\end{figure*}

\section{BSGs as extragalactic distance indicators -- the FGLR-method}

As first demonstrated by \citet{kud03} there is a very simple and compelling way 
to use BSGs for distance determinations. Massive stars with masses in the range from 12 to 
40 M$_\odot$ evolve through the BSG stage at roughly constant luminosity. In addition, since the 
evolutionary timescale is very short when crossing through the BSG domain, the amount of 
mass lost in this stage is small. As a consequence, the evolution proceeds at constant mass 
and constant luminosity. This has a very simple, but very important consequence for the 
relationship between gravity and effective temperature along the evolution towards the 
RSG stage, namely that the flux-weighted gravity log g$_{F}$, defined as log~g$_{F}$ = log~g - 4log (T$_{\rm eff}$/10$^{4}$),
stays constant. As shown 
in detail by Kudritzki et al. (2008) this immediately leads to the 
flux-weighted gravity--luminosity relationship (FGLR):

\begin{equation}
    M_{\rm bol}\,=\,a (\log\,g_F\,-\,1.5)\,+\,b
  \end{equation}

In practice this means that, after careful calibration the 
luminosity of BSGs can be inferred  by a purely spectroscopic method from measurements of 
the effective temperature and effective gravity alone. Using the new spectral diagnostic 
techniques described above, \citet{kud08} determined blue supergiant temperatures 
and gravities for a large sample of BSGs in NGC~300.  They then used the comparison of the 
calculated spectral energy distributions with multi-color HST photometry to precisely 
determine interstellar reddening and extinction, in order to obtain de-reddened visual 
and bolometric magnitudes. This revealed a beautiful and tight FGLR (Fig.\,\ref{fig13}). Including the results 
from quantitative spectroscopy of eight more galaxies then led to a first calibration (Fig.\,\ref{fig14} ).  
With a relatively small residual scatter of $\approx$ 0.3 mag the observed FGLR is an excellent tool to 
determine accurate spectroscopic distance to galaxies. It requires multicolor photometry and low 
resolution (5\AA) spectroscopy to determine effective temperature and gravity and, thus, 
flux-weighed gravity directly from the spectrum. With effective temperature, gravity
and metallicity determined one also knows the bolometric correction, which is small for A 
supergiants, which means that errors in the stellar parameters do not largely affect the 
determination of bolometric magnitudes. Moreover, one knows the intrinsic stellar SED and,
therefore, can determine interstellar reddening and extinction from the multicolor photometry, 
which then allows for the accurate determination of the reddening-free apparent bolometric 
magnitude. The application of the FGLR then yields absolute magnitudes and, thus, the distance modulus.

\begin{figure}

\begin{center}
 \includegraphics[width=8cm]{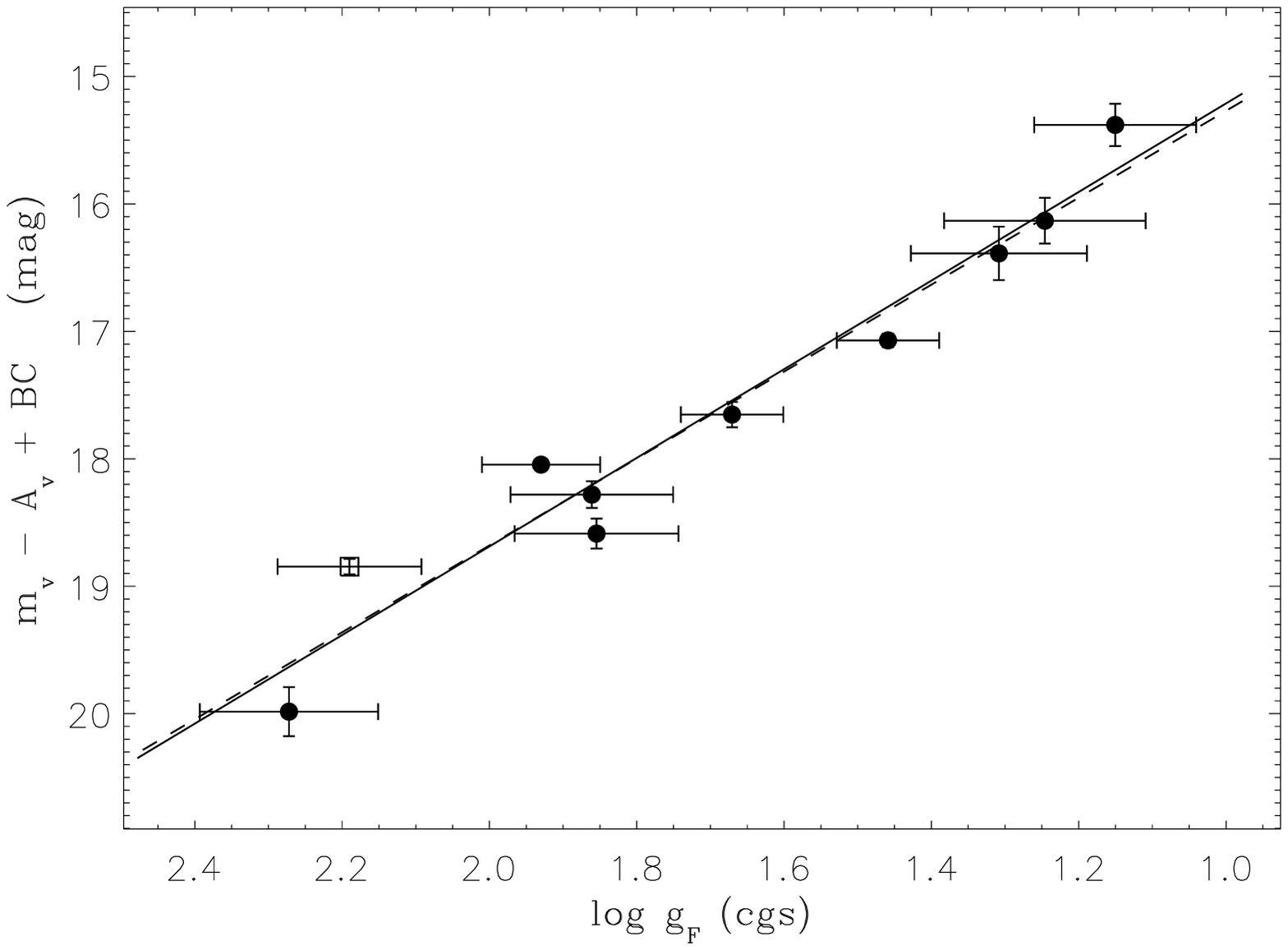} 

\caption{\small The FGLR of the Local Group dwarf irregular galaxy WLM, 
based on apparent bolometric magnitudes ($m_{bol}= m_v- A_v + BC$). The 
solid line corresponds to the FGLR calibration. The distance is determined 
by using this calibration through a minimization of the residuals. From 
\cite{urbaneja08}. 
}
   \label{fig15}
\end{center}
\end{figure}

\begin{figure}

\begin{center}
 \includegraphics[width=6cm, angle=90]{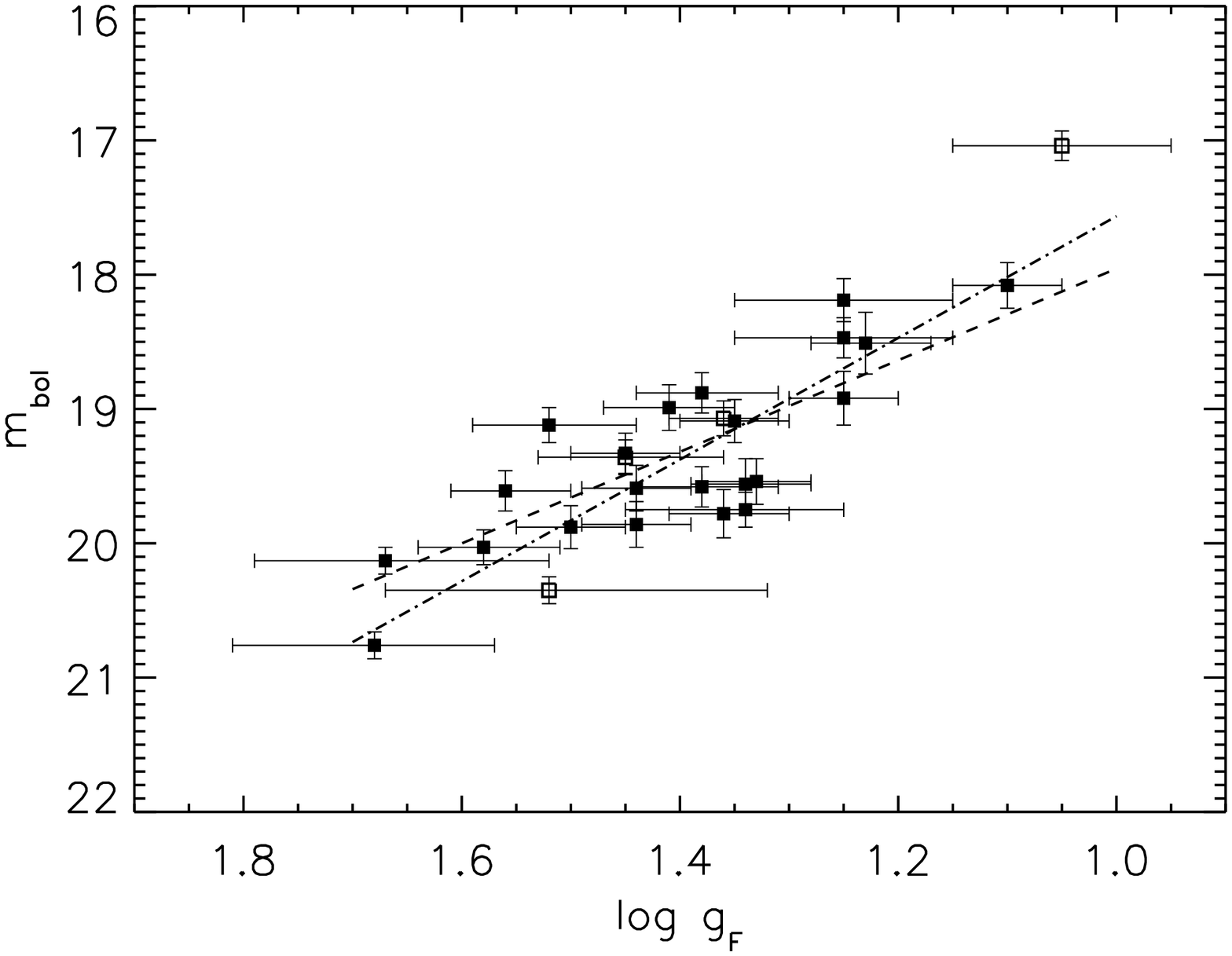} 

\caption{\small The observed FGLR in M81. Solid squares are targets used for the distance 
determination fit. The dashed line corresponds to the FGLR calibration by \citet{kud08}. 
The dashed-dotted line is a new (still preliminary) LMC calibration
\citep{urbaneja11}. Both calibrations 
yield a very similar distance modulus. From \citet{kud11}.
}
   \label{fig16}
\end{center}
\end{figure}

\section{First distance determinations with the FGLR}

The first distance determination of this type has been carried out by \citet{urbaneja08} 
who studied blue supergiants in WLM, one of the faintest dwarf irregular galaxies in the Local Group. 
The quantitative spectral analysis of VLT FORS spectra yields an extremely low metallicity of the 
young stellar population in this galaxy with an average of -0.9 dex below the solar value. 
The interstellar extinction is again extremely patchy ranging from 0.03 to 0.30 mag in E(B-V) 
(note that the foreground value given by \citealt{schlegel98} is 0.037 mag). The individually 
de-reddened FGLR - in apparent bolometric magnitude - is shown in Fig.\,\ref{fig15}  yielding a distance modulus of 
24.99 $\pm{0.10}$ mag in very good agreement with the TRGB distance \citep{rizzi07} and the K-band 
Cepheid distance \citep{gieren08}.

\citet{u09} have used KECK DEIMOS and ESI spectra to determine a distance 
to the Triangulum galaxy M33. The case of M33 is particularly interesting, since many independent 
distance determinations have been carried out for this galaxy during the last decade using a 
variety of techniques, including Cepheids, RR Lyrae, TRGB, red clump stars, planetary nebulae, 
horizontal branch stars and long-period variables. The surprising result of all of these studies has 
been that the distance moduli obtained with these different methods differ by as much as 0.6 mag, 
which is more than 30\% in linear distance. The FGLR-method yields a long distance modulus for 
M33 of 24.93 $\pm{0.11}$ mag, in basic agreement with a TRGB distance of 24.84 $\pm{0.10}$ mag obtained 
by the same authors from HST ACS imaging. The long distance modulus agrees also very well with 
the eclipsing binary distance obtained by \citet{bonanos06}.  
U et al.  relate the difference between their result and the published Cepheid distances to 
the difference in accounting for interstellar reddening (see Fig.\,\ref{fig1}).

\citet{kud11} analyzed Keck LRIS spectra of BSGs in M81 and applied the FGLR method to obtain a distance 
modulus of 27.7 $\pm{0.1}$ mag in good agreement with recent TRGB work (Fig.\,\ref{fig16}). 
The comparison with Cepheid distances is discussed in the next section.

These successful first applications of the FGLR-method for distance determinations indicate 
the very promising potential of the method to provide an independent constraint on the 
extragalactic distance scale.

\section{BSG metallicity and distance determinations and the metallicity dependence of 
the Cepheid period-luminosity relationship}

In this section we use the results from the BSG studies to discuss Cepheid distances and 
the ``standard'' approach for their metallicity correction. We start with M81 (the discussion 
presented here closely follows the one given by \citealt{kud11}). In addition to the HST Key 
Project work on M81 \citep{freedman94, freedman01} there are two recent Cepheid studies by \citet{mccommas09} 
and by \citet{gerke11}. \citet{mccommas09} use HST light curves of 11 fundamental and two first overtone 
short period Cepheids in the outer disk of M81 at $\sim$ 13.5 kpc galactocentric distance. 
\citet{gerke11} investigate 107 long period Cepheids observed with the LBT in a galactocentric range of 
3.5 to 10.5 kpc with ground-based B, V, I photometry. Applying the Wesenheit VI method, both studies obtain a distance modulus relative to the LMC which is 0.17 mag 
longer than the BSG result. However, this is based on a metallicity correction $\Delta \mu$ = $\gamma$([Z] - [Z]$_{LMC}$),
which is motivated by the fact that the inner field Cepheids observed by the KP yield a distance modulus, which is 0.23 
mag shorter relative to the outer fields without such a correction applied. They use \citet{zaritsky94} for metallicity [Z] and the metallicity gradient and
[Z]$_{LMC}$=-0.19 for the LMC oxygen abundance (see section 4, 2nd to last paragraph). This metallicity gradient requires a highly negative value of $\gamma$,
$\gamma$ = -0.55 mag dex$^{-1}$, to account for the inner and outer field Cepheid distance moduli differences.

Applying the BSG metallicity gradient would require an even more negative value of 
$\gamma$, namely $\gamma$ = -0.65 mag dex$^{-1}$. Moreover, the LMC oxygen abundance in these corrections is too large compared with the 
LMC oxygen abundance of B-stars found by \citet{hunter07} ([Z]$_{LMC}$ = -0.36 dex), the iron abundances of LMC 
Cepheids determined by \citet{romaniello08} and \citet{luck98} ([Z]$_{LMC}$ = -0.33 dex), and the LMC \hii~region oxygen abundances 
obtained by \citet{bresolin11} ([Z]$_{LMC}$ = -0.33 dex). This means that with the BSG metallicity values 
in M81 the Cepheids in the outer field have a metallicity 0.11 dex higher than the LMC. If one would apply the metallicity correction
with $\gamma$ = -0.65 mag dex$^{-1}$ accordingly, this would enlarge the distance modulus by another 0.07 mag.

However, with such a large negative value of $\gamma$ it is important to note that this empirical correction, which claims that Cepheids become brighter with 
increasing metallicity, is in striking disagreement with pulsation theory, which predicts exactly the opposite, 
namely that the Cepheid brightness decreases with increasing metallicity \citep{fiorentino02, marconi05, fiorentino07, bono08}.
It also disagrees with the recent high S/N, high spectral resolution quantitative spectroscopy in the Milky Way and the LMC 
carried out by \citet{romaniello08}, which confirms the prediction by pulsation theory. According to this work, 
the value of $\gamma$ should be positive and not negative. {\it In other words, as careful spectroscopic metallicity studies 
compared with observed differences of distance moduli between inner and outer field Cepheids push $\gamma$ to increasingly
negative values, an explanation of that distance modulus differences in terms of metallicity seems unlikely. It must be 
something else and it is an additional systematic effect not understood.} 

We also note that \citet{u09} have demonstrated from their quantitative spectroscopy of blue supergiants in M33  
that the difference of distance moduli between inner field and outer field Cepheids found by \citet{scowcroft09} 
would require a $\gamma$-value of -0.55 mag dex$^{-1}$. Even worse, \cite{bresolin10} re-determined \hii~region abundances in M33
using auroral lines and applying their abundance gradient to the Cepheid fields in M33 yields  $\gamma$ = - 1.2 mag dex$^{-1}$ 
(see discussion in \citealt*{bresolin11}).

Another galaxy where the comparison of Cepheids in the inner and outer fields leads to a significantly different 
distance modulus is the maser galaxy NGC 4258. This galaxy is of particular importance, since 
it has been used as the new anchor point for the extragalactic distance scale by \citet{riess09a, riess09b, riess11}  because of its 
accurately known distance from the Keplerian motion of water masers orbiting the central black hole \citep{humphreys08}. 
\citet{macri06} based on the \hii~region strong line method 
oxygen abundances by \citet{zaritsky94} derived a $\gamma$-value of -0.29 mag dex$^{-1}$. However, most recently, \cite{bresolin11} 
re-determined the \hii~region metallicities in this galaxy including the observation of auroral lines in a few cases. 
This led to a downward substantial revision of the metallicity, which seems to be close to the LMC and not strongly 
super-solar, and a very shallow abundance gradient. Based on these results, \cite{bresolin11} show that 
$\gamma$ = - 0.69 mag dex$^{-1}$ would be needed to explain the distance modulus difference between inner and outer fields, 
again a value much too negative, when compared with pulsation theory and observational work on Milky Way and LMC 
Cepheids. While the improved \hii~region work on this important galaxy still awaits an independent confirmation through 
a study of BSGs, it is an additional clear indication of a systematic effect on Cepheid distance moduli not understood at this point.
\citet{majaess11} discuss the large metallicity corrections suggested by \citet{gerke11} and by the recent HST/ACS Cepheid 
study of M101 by \citet{shappee11} and demonstrate that such corrections lead to very improbable distances of the LMC and SMC. 
The work by \citet{storm11} indicates that a lower limit for $\gamma$ is -0.2 mag dex$^{-1}$ 
(see also the contribution by Jesper Storm in this volume). \citet{majaess11} 
argue that crowding is very likely responsible for the distance modulus differences obtained between inner and outer 
field Cepheids and not metallicity. We think that a careful spectroscopic investigation of galactic metallicities and 
their gradients and distance determinations using the FGLR as an independent method will help to clarify the situation.

\section{The potential of BSG spectroscopy}

In this paper we have demonstrated that the quantitative spectroscopy of BSGs is a promising tool to constrain the 
chemical evolution of galaxies and to determine their distances through the FGLR-method. 
The BSG results with regard to metallicity and metallicity gradient confirmed 
previous studies based on the use of auroral lines of \hii~regions that the systematic differences between 
distance moduli obtained from inner and outer field 
Cepheids (found in M33, M81, M101, NGC 4258) are very likely not caused by a metallicity dependence of the 
period-luminosity relationship of Cepheids. There must be another reason for these systematic differences.

The FGLR technique has the unique advantage that individual reddening and extinction values, 
together with metallicity, can be determined for each supergiant target directly from spectroscopy 
combined with photometry. This reduces significantly the systematic uncertainties affecting competing 
methods such as Cepheids or TRGB. The results by \citet{kud08,kud11} and \citet{u09} for NGC 300, M 81 and M33, respectively, 
demonstrate that the reddening in spiral galaxies is extremely patchy so that the application of an \textquoteleft 
average extinction\textquoteright~correction, even with the correct value, is questionable. 
The FGLR technique is robust. \citet{bresolin04, bresolin06} have shown, 
from observations in NGC 300 and WLM, that the photometric variability of BSGs has negligible effect on the distances determined 
through the FGLR. Moreover, the study by \citet{bresolin05} also confirms that with HST photometry the FGLR method is not 
affected by crowding out to distances of at least 30 Mpc. This is the consequence of the enormous intrinsic brightness of 
these objects, which are 3 to 6 magnitudes brighter than Cepheids. It is evident that the type of work described in this 
paper can be in a straightforward way extended to the many spiral galaxies and star forming galaxies in a surprisingly 
large volume the local universe. By using present day 8m to 10m class telescopes and the existing very efficient multi-object 
spectrographs one can reach down with sufficient S/N to V = 22.5 mag in two nights of observing time under very 
good conditions. For objects brighter than M$_{V}$ = -8 mag this means metallicities and distances can be determined 
out to distances of 12 Mpc ($\mu$ = 30.5mag). With the next generation of extremely large telescopes such as the 
TMT, GMT or the E-ELT the limiting magnitude can be pushed to V = 24.5 equivalent to distances of 30 Mpc ($\mu$ = 32.5 mag).

The application of the FGLR-method and the metallicity diagnostic on a large sample of galaxies will provide an independent 
check of the systematics of the well established methods, such as the use of Cepheids and the TRGB. In this way we will be 
able to investigate the effects of metallicity and extinction, which continue to be important issues. One first crucial step 
will be the spectroscopic investigation of the population of BSGs in the maser galaxy NGC~4258, the very promising new anchor 
point of the extragalactic distance scale. This work is under way.

One weakness of the FGLR-method at this stage is that its calibration relies on the assumption of a distance to the LMC. 
No reliable parallaxes are available for a sufficiently large sample of BSGs in the MW. However, this will change with 
GAIA mission, which will deliver accurate distances to hundreds of MW BSGs allowing for a precise galactic calibration of the FGLR. 
With this calibration BSGs will become powerful primary distance indicators enabling a very accurate determination of the Hubble constant.

\acknowledgments
This work is the result of an ongoing collaborative effort over many years. We wish to thank
the ``Araucarians'' Wolfgang Gieren and Grzegorz Pietrzynski for inviting us to join their project and for their many contributions.
Norbert Przybilla from Bamberg Observatory and Joachim Puls from Munich University Observatory have developed and provided the model 
atmosphere and radiative transfer framework, which is essential for our work. Special thanks go to our colleagues in Hawaii, Fabio Bresolin, Zach Gazak, 
and Vivian U for their tremendous dedication and skillful contributions to make this project happen.

This work was supported by the National Science Foundation under grant AST-1008798. In addition,
RPK acknowledges support by the Alexander-von-Humboldt Foundation and the hospitality 
of the Max-Planck-Institute for Astrophysics in Garching and the University Observatory Munich.



\end{document}